\documentclass[10pt,conference]{IEEEtran}
\pdfoutput=1
\usepackage{hyperref}
\usepackage{cite}
\usepackage{amsmath,amssymb,amsfonts}
\usepackage{algorithmic}
\usepackage{graphicx}
\usepackage[caption=false]{subfig}
\usepackage{textcomp}
\usepackage{xcolor}
\usepackage{color, soul}
\usepackage{enumitem}
\usepackage{import}
\usepackage[ruled,vlined]{algorithm2e}
\usepackage{setspace}

\usepackage{tikz}
\newcommand*\circled[1]{\tikz[baseline=(char.base)]{
            \node[shape=circle,draw,inner sep=1pt] (char) {#1};}}
\def\BibTeX{{\rm B\kern-.05em{\sc i\kern-.025em b}\kern-.08em
    T\kern-.1667em\lower.7ex\hbox{E}\kern-.125emX}}
\begin{document}

\title{Harness the Power of DERs for Secure Communications in Electric Energy Systems}

\author{\IEEEauthorblockN{\textbf{Ioannis Zografopoulos, Juan Ospina, Charalambos Konstantinou}}
\IEEEauthorblockA{Department of Electrical and Computer Engineering, FAMU-FSU College of Engineering \\
Center for Advanced Power Systems, Florida State University\\
Email:\{izografopoulos, jjospina, ckonstantinou\}@fsu.edu}}

\maketitle

\begin{abstract}
Electric energy systems are undergoing significant changes to improve system reliability and accommodate increasing power demands. The penetration of distributed energy resources (DERs) including roof-top solar panels, energy storage, electric vehicles, etc., enables the on-site generation of economically dispatchable power curtailing operational costs. The effective control of DERs requires communication between utilities and DER system operators. The communication protocols employed for DER management and control lack sophisticated cybersecurity features and can compromise power systems secure operation if malicious control commands are issued to DERs. To overcome authentication-related protocol issues, we present a bolt-on security extension that can be implemented on Distributed Network Protocol v3 (DNP3). We port an authentication framework, \textit{DERauth}, into DNP3, and utilize real-time measurements from a simulated DER battery energy storage system to enhance communication security. We evaluate our framework in a testbed setup using DNP3 master and outstation devices performing secure authentication by leveraging the entropy of DERs.

\end{abstract}


\section{Introduction} \label{s:introduction}

As the electric power grid advances towards a decentralized architecture, distributed energy resources (DERs) are becoming more prevalent all over the globe. DER generation capacity is expected to proliferate from $132.4$ GW in 2017 to $528.4$ GWs by 2026 due to their contribution in more efficient grid operation, economical power, low carbon emissions, and more reliable power systems~\cite{DERgoals, DERvalue}. DERs are small-scale power generation or storage assets, often in the range of 1 kW to 10 MW, placed on the distribution level close to consumers and loads. 
Examples of DERs include PV systems, electric vehicles, wind turbines, etc. often bolstered by energy storage solutions such as battery energy storage systems (BESS), fuel cells, and flywheels. BESS-based DERs incorporate mostly lithium-ion (li-ion) battery cells as their energy storage elements due to their long life-cycles, high energy density, and low maintenance costs~\cite{Batt_storage}.

To harness the power of DERs for grid services, interconnection and interoperability standards, such as  IEEE 1547-2020, enforces DERs to include communication interfaces enabling local and bulk power system management and control~\cite{IEEE1547}. DER-to-utility or DER-to-aggregator communication is supported by embedded devices and network protocols~\cite{muyeen2017communication}. Prominent communication protocols include IEEE 1815-Distributed Network Protocol v3 (DNP3), Modbus, and IEEE 2030.5~\cite{Johnson2017Roadmap}. Their non-proprietary nature aids manufacturers in developing custom implementations which lack robust security features or require computationally intensive processes to meet security requirements~\cite{jin2011event, carter2017cyber}.

Although IEEE 1547-2020 Std. mandates secure communications for DER monitoring and control, there is a plethora of already deployed legacy devices utilizing insecure or outdated communication protocol versions. For example, more than $75$\% of North American utilities employ DNP3 for supervisory control and data acquisition~\cite{jin2011event}. 
DNP3 vulnerabilities undermine not only the security of the communication channel but also the security of devices relying on this protocol. In~\cite{DNP3_taxonomy} a comprehensive presentation of DNP3 vulnerabilities is presented. Recently, DNP3 received updates to conform with IEEE 1547-2020 and feature cryptographically secure properties similar to newer protocols (e.g., IEEE 2030.5)~\cite{IEEE1815}, however, the protocol has still significant drawbacks and legacy devices still remain a valid concern~\cite{DNP3-SA, mclaughlin2016cybersecurity}.

As with most cybersecurity problems, there is no single silver bullet which can single-handedly fend off all attacks~\cite{stright2020defensive}. We provide a security extension that can enhance utility-to-DER communications while requiring minimum redesign efforts. In this work, we leverage the inherent cell entropy from BESS-enabled DER devices, demonstrated in DERauth~\cite{zografopoulos2020derauth}, to develop a hardware-based security primitive which can be incorporated in DNP3 communications and support secure authentication. We port DERauth functionality to the OpenDNP3 library, an open source reference implementation for the DNP3 protocol~\cite{opendnp3ref}. DERauth's security mechanism leverages a challenge-reply scheme, which incorporates real-time BESS measurements, to generate unique cell signatures (secure authentication). Experimental results implemented on the OpenDNP3 stack validate our scheme using battery data generated by simulated li-ion cells. Specifically, our contributions can be summarized as follows: \textit{(i)} we expand our previous work and port DERauth's BESS-based authentication scheme into the OpenDNP3 protocol stack, \textit{(ii)} we utilize simulated li-ion battery measurements factoring their real-world discrepancies for the secure authentication process, and \textit{(iii)} we provide the implementation details of the design which can be ported in any DNP3-assisted industrial control environment reinforcing communications security. 

The rest of the paper is organized as follows. Section II describes our methodology. Section III presents the experimental setup and results of our implementation. Section IV concludes the paper and discusses future work.

\section{{DERauth} Framework}  \label{s:methodology}

\begin{figure}[t]
\centerline{\includegraphics[width=0.42\textwidth]{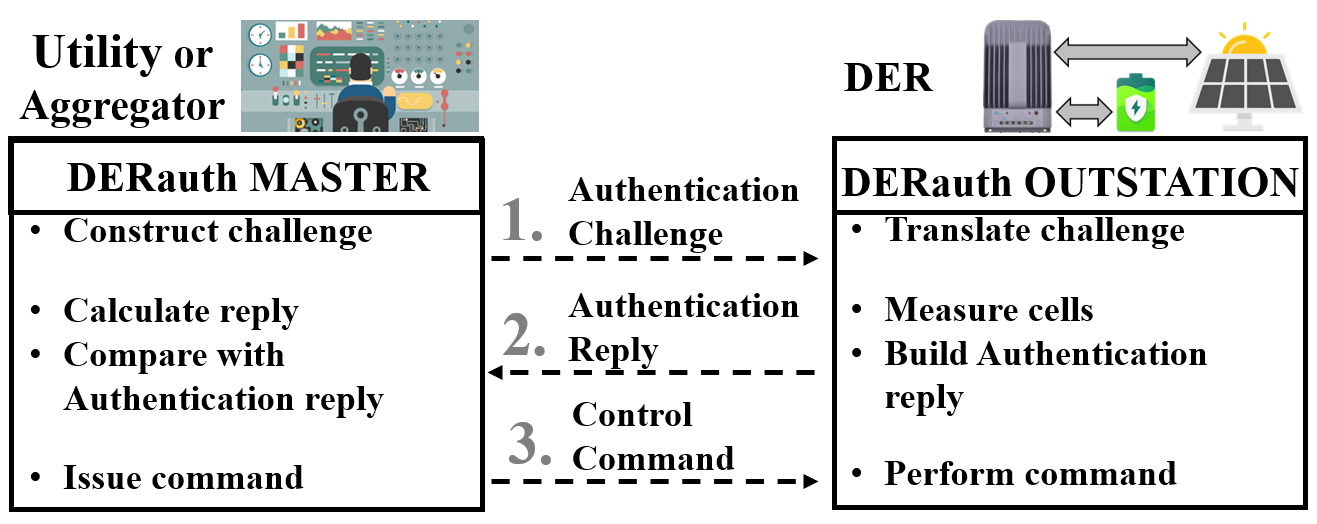}}
\caption{DERauth secure authentication framework.}
\label{fig:DERauth}
\vspace{-4mm}
\end{figure}

The DERauth framework secures the measurement and control command data issued to DERs. It provides a security extension that can be retrofitted to any BESS-enabled grid asset which uses DNP3 with minimum redesign requirements, serving as an authentication layer on top of the DNP3 protocol stack. A high-level overview of the DERauth framework is presented in Fig.~\ref{fig:DERauth}. It operates using a $64$-bit \emph{challenge-reply scheme} between DNP3 master and DNP3 outstation devices (located on the DER side). It requires DER outstation devices to be authenticated before control commands can be issued from the DNP3 master. The authentication process leverages the intrinsic characteristics of the BESS li-ion battery cells to generate entropy. The concept leverages the BESS cell voltage and state-of-charge (SoC) measurements, and incorporates the inherent randomness -- caused by the unique li-ion physical properties due to their manufacturing process -- for the protocol replies. Below, we outline the core components of DERauth; more details about the implementation are provided in~\cite{zografopoulos2020derauth}. 

\subsection{System Components} \label{s:components}

The communication between a DNP3 master and DER outstation devices is initiated after the enrollment and authentication phases. 

\textit{{Enrollment phase}}: 
In this phase, the DNP3 master and DER outstation device start their communication and exchange their cell-reply tables, $C_{rt}$. The $C_{rt}$ table includes 8-bit sequences ($r_i$) for each BESS cell which are returned if the specific cell gets authenticated. $C_{rt}$ is  defined as $C_{rt}=N\times<c_i, r_i>$, where $N$ is the number of BESS cells, and $c_i$ and $r_i$ represent the $i$-th BESS cell and its  predefined reply, respectively. Notably, the contents of the $C_{rt}$ table get updated at every authentication round and at both communication ends asynchronously. Thus, in the event of a compromise, the attained information does not provide useful details to man-in-the-middle adversaries. 

\textit{{Authentication phase}}: 
This phase comprises the main portion of DERauth and can be segmented into three parts.

\circled{1} First, the DNP3 master constructs the $64$-bit challenge. 
The first $16$ bits of the $64$-bit challenge define which BESS cells will be polled for their real-time voltages and SoCs. The next $32$ bits, indicate which BESS cells need to be authenticated (their corresponding  $r_i$s from the $C_{rt}$ table will be returned if the authentication succeeds). The remaining $16$ bits instruct the DNP3 outstation device, monitoring and controlling the operation of the DER asset, regarding the type of transformation that will take place before the reply is forwarded to the DNP3 master device.

\circled{2} The next part of the authentication process occurs on the DNP3 outstation device. Upon receiving the $64$-bit challenge, the outstation first translates the challenge and builds a corresponding $64$-bit reply following the format specified by the DNP3 master. The DNP3 reply includes the real-time voltage and SoC BESS cell measurements, and  the $r_i$s from the requested cells. After this temporary reply is built, it gets transformed using a predefined transformation function, and the last $16$ bits of the issued challenge. This transformation increases the reply entropy and once the temporary reply is transformed, it is forwarded to the DNP3 master for validation. The real-time voltage and SoC measurements, in addition to the constantly updating $r_i$s and the transformation function, effectively increase the challenge-reply entropy and the security of our protocol, rendering reconstructing or decoding the exchanged data infeasible.

\circled{3} Upon receiving the $64$-bit reply, the DNP3 master calculates locally the \emph{expected} reply according to the challenge format and then compares it with the received one. If the local version of the reply -- calculated by the DNP3 master --  coincides with the received one from the DNP3 outstation, then the authentication round is performed successfully. In any other case, the round has failed, and hence, any exchanged information is dropped and the authentication phase is repeated. Only if the authentication succeeds, then both the DNP3 master and outstation can update their $C_{rt}$ table copies using the exchanged real-time voltage and SoC information. Following this procedure, static information is not retained anywhere in our system (both $C_{rt}$ tables  are updated asynchronously), impeding attackers from gaining sufficient system information and reverse-engineering our security protocol extension.

\section{Experimental Setup \& Results} \label{s:Implementation And Evaluation}
\begin{figure}[t]
\centerline{\includegraphics[width=0.45\textwidth]{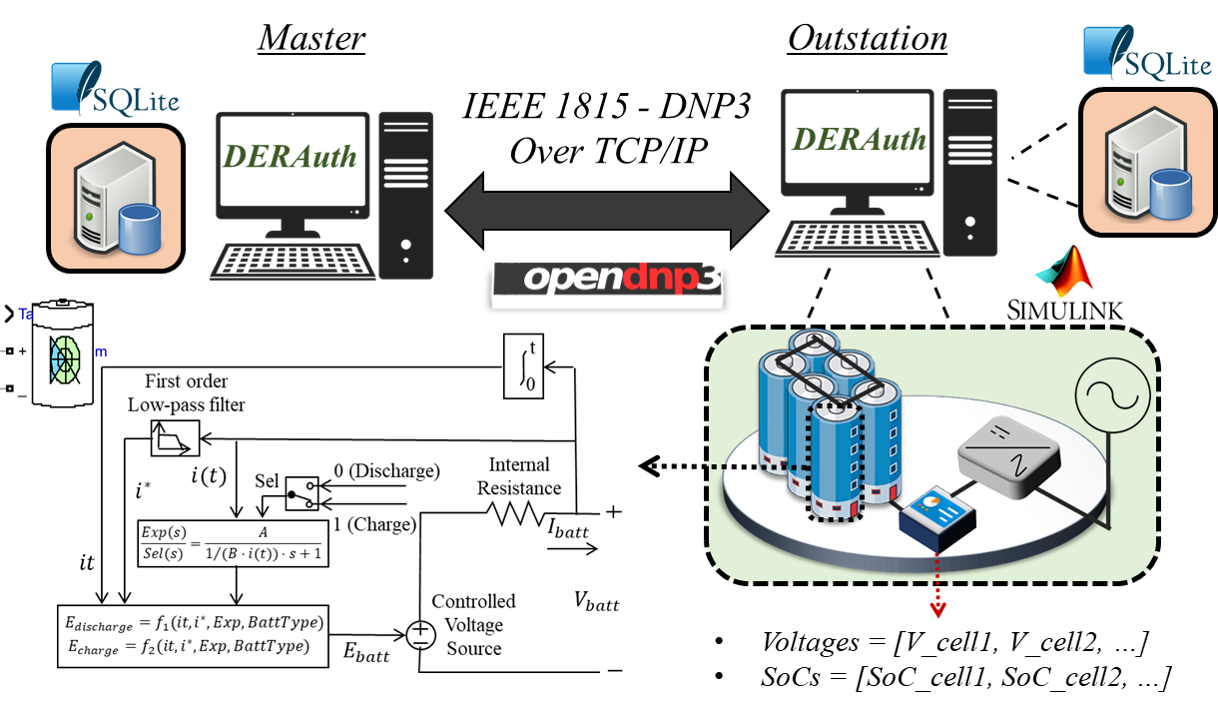}}
\caption{Top-level diagram of test setup for evaluating the DERauth process using the DNP3 communication protocol.}
\label{fig:overall_diagram}
\vspace{-4mm}
\end{figure}

\subsection{Setup \& Design Parameters} \label{s:setup}
For the evaluation of the DERauth authentication process, we implement a testbed setup where two computing stations communicate via DNP3, as master and outstation. Fig.~\ref{fig:overall_diagram} depicts the overall top-level diagram of the test setup. The master and outstation devices are running modified versions of DNP3, using the open-source OpenDNP3 library~\cite{opendnp3ref}, which integrates the DERauth challenge-reply mechanism and supports an SQLite database to store the temporary challenge-reply values exchanged during authentication. In this implementation, the DNP3 outstation is modeled as an agent that interacts with the BESS, i.e., the DER, and the master DNP3 device is modeled as an agent that sends the corresponding charge/discharge setpoints to the outstation DER.

\begin{table}[t]
\centering
\footnotesize
\setlength{\tabcolsep}{3.5pt}
\caption{Li-ion cells parameters and characteristics.}
\vspace{-2mm}
\label{tab:cells_characs}
\begin{tabular}{||c|c|c||}
\hline \hline
\textbf{Parameters} & \textbf{Cell \#1} & \textbf{Cell \#2} \\ \hline \hline
Nominal Voltage (V) & $3.5$ & $3.5$ \\ \hline
Rated Capacity (Ah) & $2$ & $2$ \\ \hline
Initial SoC (\%) & $64$ & $65$ \\ \hline
Response Time (s) & $5$ & $5$ \\ \hline
Max. Capacity (Ah) & $2.05$ & $2.02$ \\ \hline
Cut-off Voltage & $2.625$ & $2.622$ \\ \hline
Fully Charged Voltage (V) & $4.1$ & $4.1$ \\ \hline
Nominal Discharge Current (A) & $0.4$ & $0.4$ \\ \hline
Internal Resistance (Ohms) & $0.017$ & $0.012$ \\ \hline
Capacity at Nominal Voltage (Ah) & $1.8087$ & $1.7897$ \\ \hline
Exponential Zone {[}V, Ah{]} & {[}$3.88$, $0.2${]} & \multicolumn{1}{l|}{{[}$3.81$, $0.2${]}} \\ \hline
 \hline
\end{tabular}
\vspace{-1em}
\end{table}

\normalsize
\begin{figure}[t]
\centerline{\includegraphics[width=0.45\textwidth]{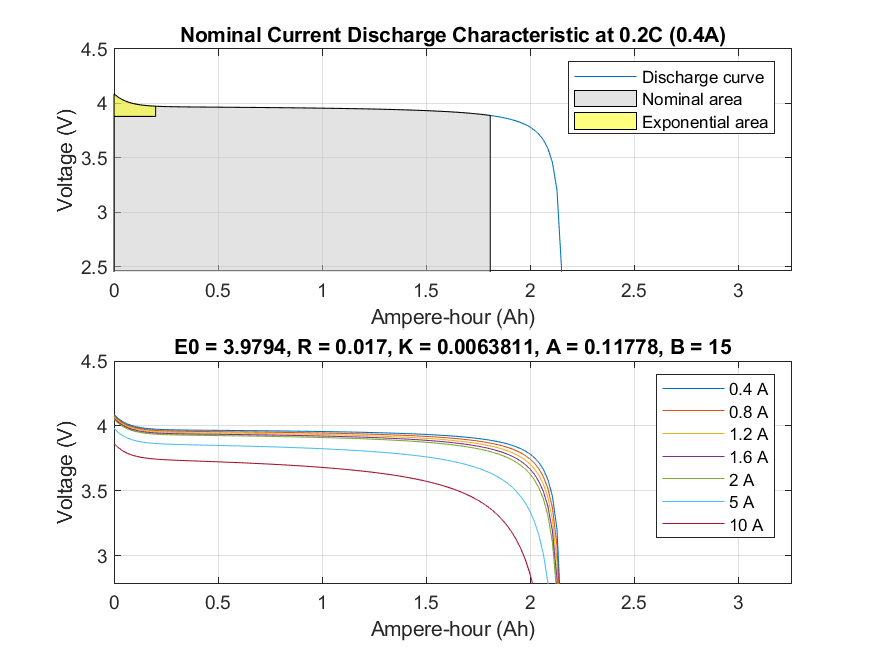}}
\vspace{-2mm}
\caption{Li-ion cell \#1 discharge characteristics. $C$ is a measure of the rate at which the battery cell is discharged relative to its max. capacity. $R$ is the internal resistance of the battery cell. The parameters $E_0$, $K$, $A$, and $B$ are parameters that model the behavior of the battery cell. $E_0$ is the constant voltage in V, $K$ is the polarization constant in V/Ah, $A$ is the exponential voltage in V, and $B$ is the exponential capacity in Ah$^{-1}$.}
\label{fig:cell1_char}
\vspace{-4mm}
\end{figure}

\begin{figure}[t]
\centerline{\includegraphics[width=0.45\textwidth]{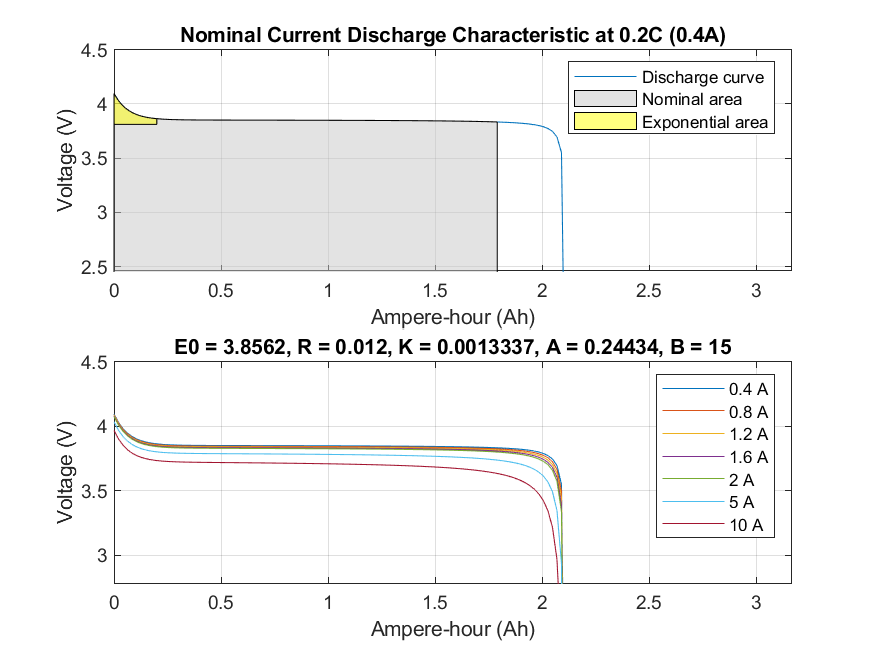}}
\vspace{-2mm}
\caption{Li-ion cell \#2 discharge characteristics. $C$ is a measure of the rate at which the battery cell is discharged relative to its max. capacity. $R$ is the internal resistance of the battery cell. The parameters $E_0$, $K$, $A$, and $B$ are parameters that model the behavior of the battery cell. $E_0$ is the constant voltage in V, $K$ is the polarization constant in V/Ah, $A$ is the exponential voltage in V, and $B$ is the exponential capacity in Ah$^{-1}$.}
\label{fig:cell2_char}
\vspace{-4mm}
\end{figure}

\subsection{DER Li-ion Battery Modeling}
The li-ion cells of the BESS interacting with the DNP3 outstation agent are modeled using the MATLAB/Simscape battery model~\cite{tremblay2009experimental}. This model provides predetermined behaviors for different types of battery chemistries such as nickel-cadmium, lead-acid, and li-ion batteries. It also supports the simulation of temperature and aging effects in the battery cells.In order to obtain realistic values from the battery cell simulations, the cell temperature is also considered, since it can affect the voltage during charge and discharge operations.

Six li-ion cells are modeled and two of those cells are used for the DERauth authentication. Figs. \ref{fig:cell1_char} and \ref{fig:cell2_char} depict the cell and discharge characteristics of the two battery cells selected to provide their respective voltages and SoC values. A battery management system is  devised to provide the voltage and SoC measurements from individual cells in the battery system. Table~\ref{tab:cells_characs} shows the parameters and characteristics that define the behavior of the battery cell models.  

\begin{figure*}[ht]
\centering
  \noindent\makebox[\textwidth]{\includegraphics[width=18.cm]{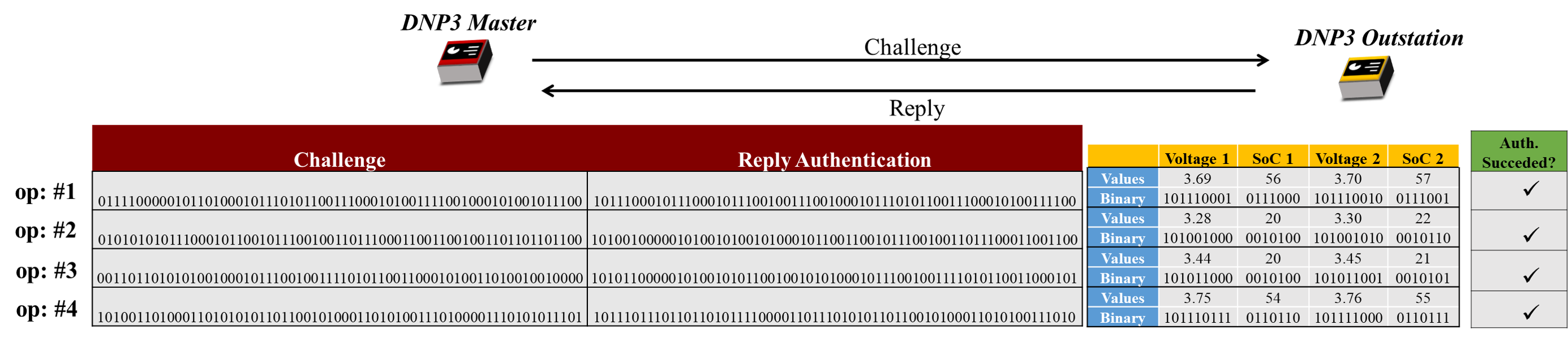}}
  \vspace{-6mm}
  \caption{$64$-bit challenge and replies exchanged between the DNP3 master and outstation during the DERauth authentication process.
  }
  \label{fig:auth_results}
\vspace{-4mm}
\end{figure*}

\begin{figure}[t]
\centerline{\includegraphics[width=0.32\textwidth]{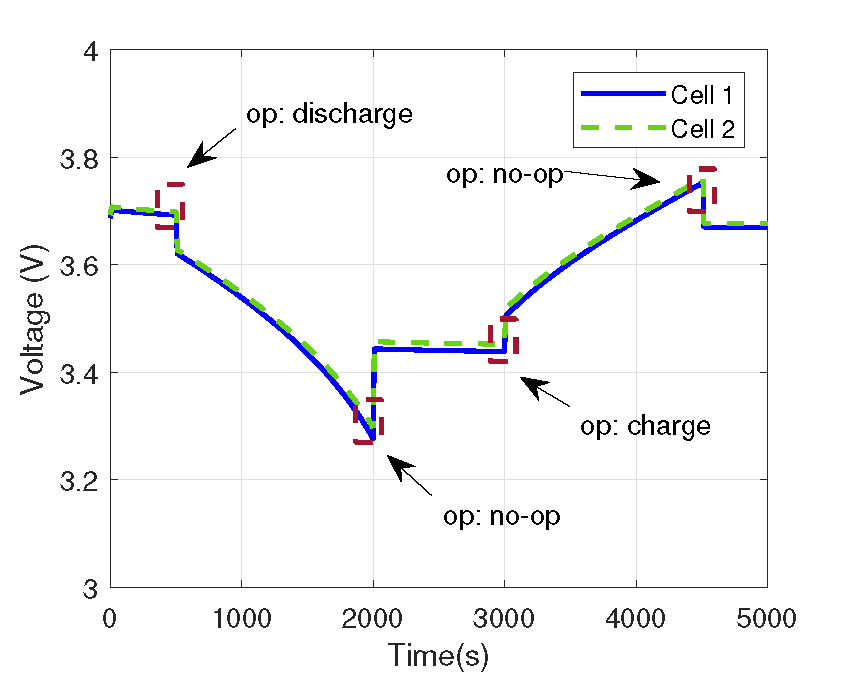}}
\vspace{-2mm}
\caption{Voltage measurements for li-ion cells during test scenario operations. The red boxes mark the time when DERauth is performed $5$ \textit{seconds} before an operation is carried out.}
\label{fig:vol_cells}
\vspace{-4mm}
\end{figure}

\begin{figure}[t]
\centerline{\includegraphics[width=0.32\textwidth]{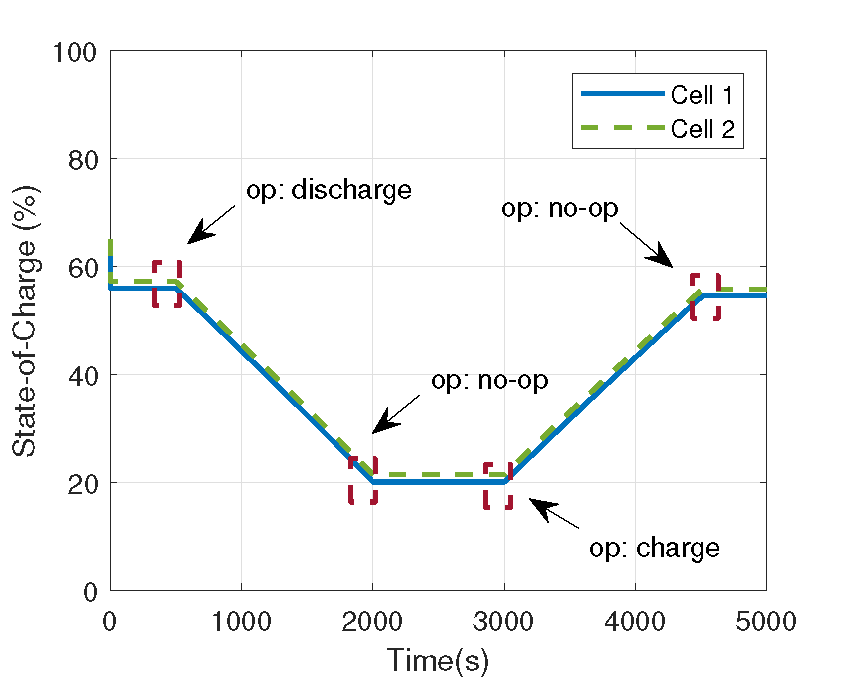}}
\vspace{-2mm}
\caption{State-of-charge (SoC) measurements for li-ion cells during test scenario operations. The red boxes mark the time when DERauth is performed $5$ \textit{seconds} before an operation is carried out.}
\label{fig:soc_cells}
\vspace{-4mm}
\end{figure}

\vspace{-1mm}

\subsection{Case Study \& Results}
To assess the operation of DERauth within the test setup presented in Section~\ref{s:setup}, a modified version of OpenDNP3 integrating DERauth is deployed in two different computers connected via an Ethernet switch. The DNP3 communication protocol runs over TCP/IP and the master and outstation are configured to interchange \textit{$64$-bit float Analog} values, i.e., \textit{Group $32$, Variation $8$}~\cite{IEEE1815}. 

In this case study, the master device initiates the DERauth authentication process every time it sends a new setpoint command to the outstation controlling the DER.  Specifically, in our scenario, the DERauth authentication challenge is predefined to occur $5$ \textit{seconds} before the master sends the corresponding operation command to the respective outstation. Although a $5$-second interval is selected for this case study,  any other value could also be used depending on the operational requirements. Figs.~\ref{fig:vol_cells} and~\ref{fig:soc_cells} present the voltage and SoC measurements from the two cells used in DERauth. The total simulation time is $5000$ \textit{seconds}, and four battery operations are requested by the master controller. These operations are: \textit{(i)} a $10$ W discharge request at $500$ \textit{seconds}, \textit{(ii)} a no-operation (\texttt{no-op}) request at $2000$ \textit{seconds}, \textit{(iii)} a $10$ W charge request at $3000$ \textit{seconds}, and \textit{(iv)} a no-operation (\texttt{no-op}) request at $4500$ \textit{seconds}. At each of these points, DERauth is executed  before ($5$ seconds) the setpoint command is issued. The specific values exchanged by the master and outstation devices during these four authentications are presented in Fig.~\ref{fig:auth_results}. Notably, the voltage and SoC values polled, are used to construct the reply for each corresponding challenge.  

\section{Conclusion and Future work}\label{s:Conclusion}

This paper demonstrates the implementation of a hardware-based authentication framework for DERs. We port DERauth, which leverages existing hardware and li-ion manufacturing variations, into OpenDNP3 enabling BESS-based DER system secure authentication. The implementation is evaluated in a testbed setup comprised of a DNP3 master and outstation device with a simulated li-ion BESS. Future work will focus on assessing the DERauth secure authentication extension in a full-fledged distribution system with multiple DERs.

Master devices upon verifying the authenticity of DER assets will be able to issue control commands to the respective outstation devices. Real-time simulation environments will be used to evaluate the practicality of the DERauth authentication process against other protocols implementations in realistic scenarios and under stringent operational constraints (e.g., traffic, latency)~\cite{cyberpels2020}. Finally, the necessary application program interface to integrate our hardware-based security scheme in other existing protocols, such as Modbus and  IEEE 2030.5, will be designed.

\vspace{-1mm}
\bibliographystyle{IEEEtran}

{\bibliography{biblio}}

\begin{thebibliography}{10}
\providecommand{\url}[1]{#1}
\csname url@samestyle\endcsname
\providecommand{\newblock}{\relax}
\providecommand{\bibinfo}[2]{#2}
\providecommand{\BIBentrySTDinterwordspacing}{\spaceskip=0pt\relax}
\providecommand{\BIBentryALTinterwordstretchfactor}{4}
\providecommand{\BIBentryALTinterwordspacing}{\spaceskip=\fontdimen2\font plus
\BIBentryALTinterwordstretchfactor\fontdimen3\font minus
  \fontdimen4\font\relax}
\providecommand{\BIBforeignlanguage}[2]{{%
\expandafter\ifx\csname l@#1\endcsname\relax
\typeout{** WARNING: IEEEtran.bst: No hyphenation pattern has been}%
\typeout{** loaded for the language `#1'. Using the pattern for}%
\typeout{** the default language instead.}%
\else
\language=\csname l@#1\endcsname
\fi
#2}}
\providecommand{\BIBdecl}{\relax}
\BIBdecl

\bibitem{DERgoals}
G.~Insights, ``{Global Capacity of Distributed Energy Resources Is Expected to
  Reach Nearly 530 GW in 2026},'' 2017.

\bibitem{DERvalue}
S.~Tierney, ``{The Value of “DER” to “D”: The Role of Distributed
  Energy Resources in Supporting Local Electric Distribution System
  Reliability},'' {Analysis Group, Inc.}, Tech. Rep., 2016.

\bibitem{Batt_storage}
A.~K. {Chen}, Y.~N. {Velaga}, and P.~K.~P. {Sen}, ``Future electric power grid
  and battery storage,'' in \emph{IEEE Texas Power and Energy Conference
  (TPEC)}, 2019, pp. 1--6.

\bibitem{IEEE1547}
{IEEE 1547}, ``{IEEE Standard Conformance Test Procedures for Equipment
  Interconnecting Distributed Energy Resources with Electric Power Systems and
  Associated Interfaces},''
  \url{https://standards.ieee.org/standard/1547\_1-2020.html}, 2020.

\bibitem{muyeen2017communication}
S.~Muyeen and S.~Rahman, \emph{Communication, control and security challenges
  for the smart grid}.\hskip 1em plus 0.5em minus 0.4em\relax The Institution
  of Engineering and Technology, 2017.

\bibitem{Johnson2017Roadmap}
J.~Johnson, ``Roadmap for photovoltaic cyber security,'' 2017.

\bibitem{jin2011event}
D.~Jin, D.~M. Nicol, and G.~Yan, ``An event buffer flooding attack in dnp3
  controlled scada systems,'' in \emph{Proceedings of the 2011 Winter
  Simulation Conference (WSC)}.\hskip 1em plus 0.5em minus 0.4em\relax IEEE,
  2011, pp. 2614--2626.

\bibitem{carter2017cyber}
C.~Carter \emph{et~al.}, ``Cyber security assessment of distributed energy
  resources,'' in \emph{Proceedings of the IEEE Photovoltaic Specialists
  Conference (PVSC), Washington, DC, USA}, 2017, pp. 25--30.

\bibitem{DNP3_taxonomy}
S.~East \emph{et~al.}, ``A taxonomy of attacks on the dnp3 protocol,'' in
  \emph{Critical Infrastructure Protection III}, C.~Palmer and S.~Shenoi,
  Eds.\hskip 1em plus 0.5em minus 0.4em\relax Berlin, Heidelberg: Springer
  Berlin Heidelberg, 2009, pp. 67--81.

\bibitem{IEEE1815}
1815.1-2015, ``{IEEE Standard for Exchanging Information Between Networks
  Implementing IEC 61850 and IEEE Std 1815(TM) [Distributed Network Protocol
  (DNP3)]},'' 2020.

\bibitem{DNP3-SA}
C.~Cremers, M.~Dehnel-Wild, and K.~Milner, ``{Secure authentication in the
  grid: A formal analysis of DNP3 SAv5.}'' \emph{Journal of Computer Security},
  vol.~27, no.~2, pp. 203 -- 232, 2019.

\bibitem{mclaughlin2016cybersecurity}
S.~McLaughlin \emph{et~al.}, ``The cybersecurity landscape in industrial
  control systems,'' \emph{Proceedings of the IEEE}, vol. 104, no.~5, pp.
  1039--1057, 2016.

\bibitem{stright2020defensive}
J.~Stright, P.~Cheetham, and C.~Konstantinou, ``Defensive cost-benefit analysis
  of smart grid digital functionalities,'' \emph{arXiv:2008.12843}, 2020.

\bibitem{zografopoulos2020derauth}
I.~Zografopoulos and C.~Konstantinou, ``{DERauth: A Battery-based
  Authentication Scheme for Distributed Energy Resources},'' in \emph{IEEE
  Computer Society Annual Symposium on VLSI (ISVLSI)}, 2020, pp. 560--567.

\bibitem{opendnp3ref}
\BIBentryALTinterwordspacing
{Automatak LLC}. {OpenDNP3 Github repository}. [Online]. Available:
  \url{https://github.com/dnp3/opendnp3}
\BIBentrySTDinterwordspacing

\bibitem{tremblay2009experimental}
O.~Tremblay and L.-A. Dessaint, ``Experimental validation of a battery dynamic
  model for ev applications,'' \emph{World electric vehicle journal}, vol.~3,
  no.~2, pp. 289--298, 2009.

\bibitem{cyberpels2020}
C.~Ogilvie \emph{et~al.}, ``Modeling communication networks in a real-time
  simulation environment for evaluating controls of shipboard power systems,''
  \emph{arXiv preprint arXiv:2008.10441}, 2020.

\end{thebibliography}

\end{document}